\documentclass[reprint,aps,prl,superscriptaddress,showpacs,footinbib,longbibliography,amsmath,amssymb]{revtex4-1}
\usepackage{graphicx}
\usepackage{bm}

\usepackage{color}
\usepackage[colorlinks,bookmarks=false,citecolor=darkblue,linkcolor=red,urlcolor=blue]{hyperref}

\definecolor{darkred}{rgb}{0.7,0.0,0.0}

\definecolor{darkblue}{rgb}{0,0.02,0.45}
\def\cdbl{\color{darkblue}}
\definecolor{darkgreen}{rgb}{0.02,0.45,0.0}

\definecolor{violet}{rgb}{0.8,0.2,0.6}

\newcommand{\ev}{\mathbf e}
\newcommand{\Xv}{\mathbf X}
\newcommand{\Yv}{\mathbf Y}
\newcommand{\Zv}{\mathbf Z}

\begin{document}

\title{Breakdown of magnetic order in the pressurized Kitaev iridate $\beta$-Li$_2$IrO$_3$}
\author{M. Majumder}
\thanks{These authors contributed equally}
\affiliation{Experimental Physics VI, Center for Electronic Correlations and Magnetism, University of Augsburg, 86159 Augsburg, Germany}

\author{R.S. Manna}
\thanks{These authors contributed equally}
\affiliation{Experimental Physics VI, Center for Electronic Correlations and Magnetism, University of Augsburg, 86159 Augsburg, Germany}
\affiliation{Department of Physics, IIT Tirupati, Tirupati 517506, India}

\author{G. Simutis}
\affiliation{Laboratory for Muon Spin Spectroscopy, Paul Scherrer Institut, 5232 Villigen PSI, Switzerland}

\author{J.C. Orain}
\affiliation{Laboratory for Muon Spin Spectroscopy, Paul Scherrer Institut, 5232 Villigen PSI, Switzerland}

\author{T. Dey}
\author{F.~Freund}
\author{A.~Jesche}
\affiliation{Experimental Physics VI, Center for Electronic Correlations and Magnetism, University of Augsburg, 86159 Augsburg, Germany}

\author{R. Khasanov}
\affiliation{Laboratory for Muon Spin Spectroscopy, Paul Scherrer Institut, 5232 Villigen PSI, Switzerland}

\author{P.K. Biswas}
\affiliation{ISIS Pulsed Neutron and Muon Source, STFC Rutherford Appleton Laboratory, Harwell Campus, Didcot, Oxfordshire OX11 0QX, United Kingdom}

\author{E. Bykova}
\author{N. Dubrovinskaia}
\affiliation{Laboratory of Crystallography, Material Physics and Technology at Extreme Conditions, Universit\"at Bayreuth, 95440 Bayreuth, Germany}

\author{L.S.~Dubrovinsky}
\affiliation{Bayerisches Geoinstitut, Universit\"at Bayreuth, 95440 Bayreuth, Germany}

\author{R. Yadav}
\author{L. Hozoi}
\author{S. Nishimoto}
\affiliation{Institute for Theoretical Physics, IFW Dresden, 01069 Dresden, Germany}

\author{A.A. Tsirlin}
\email{altsirlin@gmail.com}

\author{P. Gegenwart}
\affiliation{Experimental Physics VI, Center for Electronic Correlations and Magnetism, University of Augsburg, 86159 Augsburg, Germany}


\begin{abstract}
Temperature-pressure phase diagram of the Kitaev hyperhoneycomb iridate $\beta$-Li$_2$IrO$_3$ is explored using 
magnetization, thermal expansion, magnetostriction, and muon spin rotation ($\mu$SR) measurements, as well as single-crystal x-ray diffraction under pressure and \textit{ab initio} calculations. The N\'eel temperature of $\beta$-Li$_2$IrO$_3$ increases with the slope of 0.9\,K/GPa upon initial compression, but the reduction in the polarization field $H_c$ reflects a growing instability of the incommensurate order. At 1.4\,GPa, the ordered state breaks down upon a first-order transition giving way to a new ground state marked by the coexistence of dynamically correlated and frozen spins. This partial freezing in the absence of any conspicuous structural defects may indicate classical nature of the resulting pressure-induced spin liquid, an observation paralleled to the increase in the nearest-neighbor off-diagonal exchange $\Gamma$ under pressure.

\end{abstract}
\maketitle

{\cdbl\textit{Introduction.}} Quantum spin liquid is an exotic state of matter that entails highly correlated spins but evades magnetic ordering down to zero temperature~\cite{savary2017}. Kitaev model plays special role in this context, because it offers analytical solution for a quantum spin liquid and hosts fractionalized excitations having potential relevance to topological quantum computing~\cite{kitaev2006,trebst2017,winter2017,hermanns2018}. The Kitaev spin liquid on the honeycomb lattice can be gapless or gapped, depending on the interaction regime. It shows peculiarities in the dynamical structure factor~\cite{knolle2014} and Raman response~\cite{nasu2016}. Many of these features are shared by the three-dimensional (3D) version of the Kitaev model on the hyperhoneycomb and stripyhoneycomb lattices~\cite{mandal2009,smith2015,perreault2015}. One additional peculiarity in this case is that the spin-liquid phase survives to finite temperatures and undergoes a phase transition to a classical paramagnet~\cite{nasu2014,kimchi2014}. This distinguishes the Kitaev spin liquid in 3D from any other instance of quantum spin liquid, because only the former shows a thermodynamic singularity~\cite{nasu2014b}.

On the experimental side, Kitaev physics in 3D can be relevant to $\beta$- and $\gamma$-polymorphs of Li$_2$IrO$_3$~\cite{winter2017}. Both compounds are magnetically ordered at low temperatures~\cite{biffin2014,takayama2015,biffin2014b,modic2014}. Their non-coplanar incommensurate spin arrangements are driven by the Kitaev interactions~\cite{kimchi2015} in combination with other exchange terms producing the nearest-neighbor spin Hamiltonian~\cite{lee2015,lee2016,ducatman2018},
\begin{equation*}
\mathrm H = \sum_{\langle ij\rangle; \alpha,\beta\neq \gamma} [J_{ij}\mathbf S_i\mathbf S_j + K_{ij}S_i^\gamma S_j^\gamma \pm\Gamma_{ij}(S_i^\alpha S_j^\beta + S_i^\beta S_j^\alpha)].
\end{equation*}
Here, $J_{ij}$ is the Heisenberg exchange term, $K_{ij}$ is the Kitaev exchange, and $\Gamma_{ij}$ stands for the off-diagonal exchange anisotropy. These exchange parameters take different values for the $X$-, $Y$- and $Z$-type Ir--Ir bonds of the hyperhoneycomb lattice, respectively.

Long-range magnetic order in $\beta$- and $\gamma$-Li$_2$IrO$_3$ restricts access to the physics of the Kitaev model in 3D. \textit{Ab initio} studies suggest that at least in $\beta$-Li$_2$IrO$_3$ magnetic interactions may change significantly under pressure~\cite{kim2016}, which should shift the system toward the exotic spin-liquid state. Experimental information remains limited to date, indicating only a reconstruction of the electronic state of Ir$^{4+}$ below 2\,GPa~\cite{veiga2017}. Here, we map out the temperature-pressure phase diagram of $\beta$-Li$_2$IrO$_3$ and show that the magnetic order disappears abruptly upon a first-order transition around 1.4\,GPa, whereas local moments persist above this pressure and form a dynamic state, albeit hindered by partial spin freezing. We identify this state as a putative classical spin liquid in line with recent theory~\cite{rousochatzakis2017} suggesting the formation of such a correlated regime in the limit of large $\Gamma$, a trend consistent with the experimental evolution of the crystal structure and ensuing exchange couplings, which we obtain on different levels of \textit{ab initio} theory.

\begin{figure}
{\centering {\includegraphics{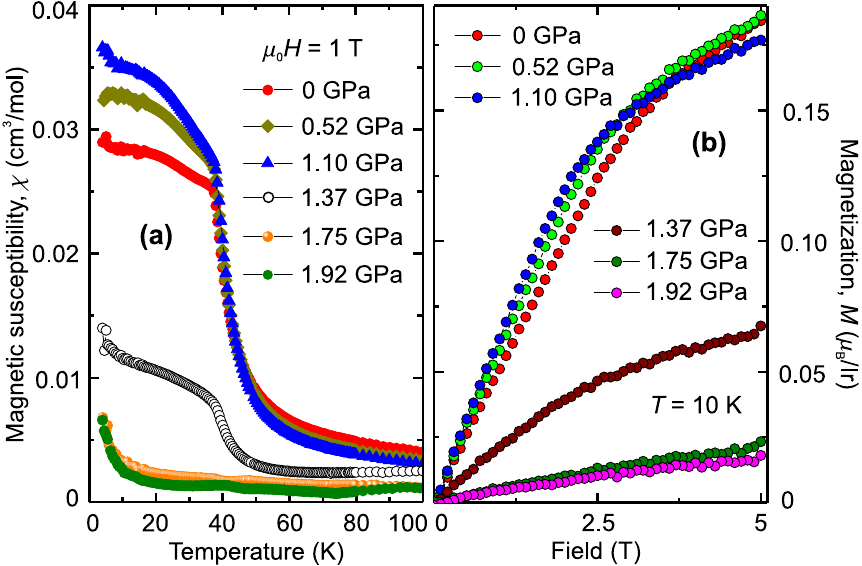}}\par} 
\caption{(a): Magnetic susceptibility ($\chi=M/H$) as a function of temperature at different pressures in the presence of the 1\,T magnetic field, (b): Magnetization curves measured at different pressures at 10\,K.} \label{fig:chi}
\end{figure}

{\cdbl\textit{Magnetization.}} 
Magnetic susceptibility was measured on polycrystalline samples of $\beta$-Li$_2$IrO$_3$. It sharply increases below 50\,K and changes slope at $T_N=38$\,K at ambient pressure (Fig.~\ref{fig:chi}a). This unusual behavior reflects the non-trivial incommensurate nature of the magnetic order~\cite{biffin2014}, which is very sensitive to the applied field. The bend around $H_c\simeq 2.7$\,T in the ambient-pressure magnetization curve (Fig.~\ref{fig:chi}b) marks the suppression of the incommensurate order by the field applied along the $b$ direction. Above 2.7\,T, commensurate spin correlations reminiscent of the zigzag order become pre-dominant~\cite{ruiz2017}. The field couples to a ferromagnetic canting mode~\cite{ruiz2017,ducatman2018}, and the value of the critical field $H_c$ gauges the stability of the incommensurate order~\cite{rousochatzakis2018}.

\begin{figure}
\includegraphics{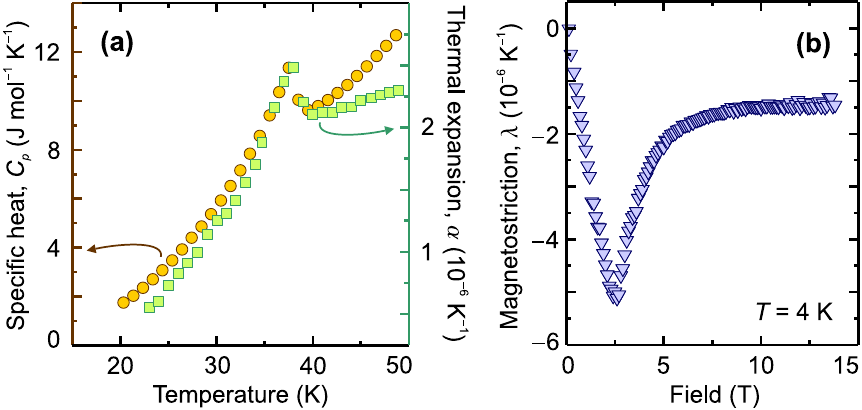}
\caption{(a): Specific heat and thermal expansion as function of temperature; (b): magnetostriction at 4 K.
} 
\label{fig:striction}
\end{figure}


Same features are seen in the magnetization data under pressure measured upon compression. Below 1.1\,GPa, $T_N$ increases with the slope of $dT_N/dp\simeq 0.9$\,K/GPa. The low-temperature susceptibility increases as well, reflecting the fact that the slope of $M(H)$ increases, and $H_c$ shifts toward lower fields. Both features are suppressed at higher pressures and no longer visible in the data collected at $1.7-1.9$\,GPa, where the signal becomes very low reaching the sensitivity limit of our measurement setup. This suppression of the magnetization is well in line with the disappearance of the x-ray magnetic circular dichroism (XMCD) signal around 1.5\,GPa~\cite{veiga2017}, because XMCD is proportional to the sample magnetization induced by the applied field. 

\begin{figure*}[tb!]
{\centering {\includegraphics[width=17cm]{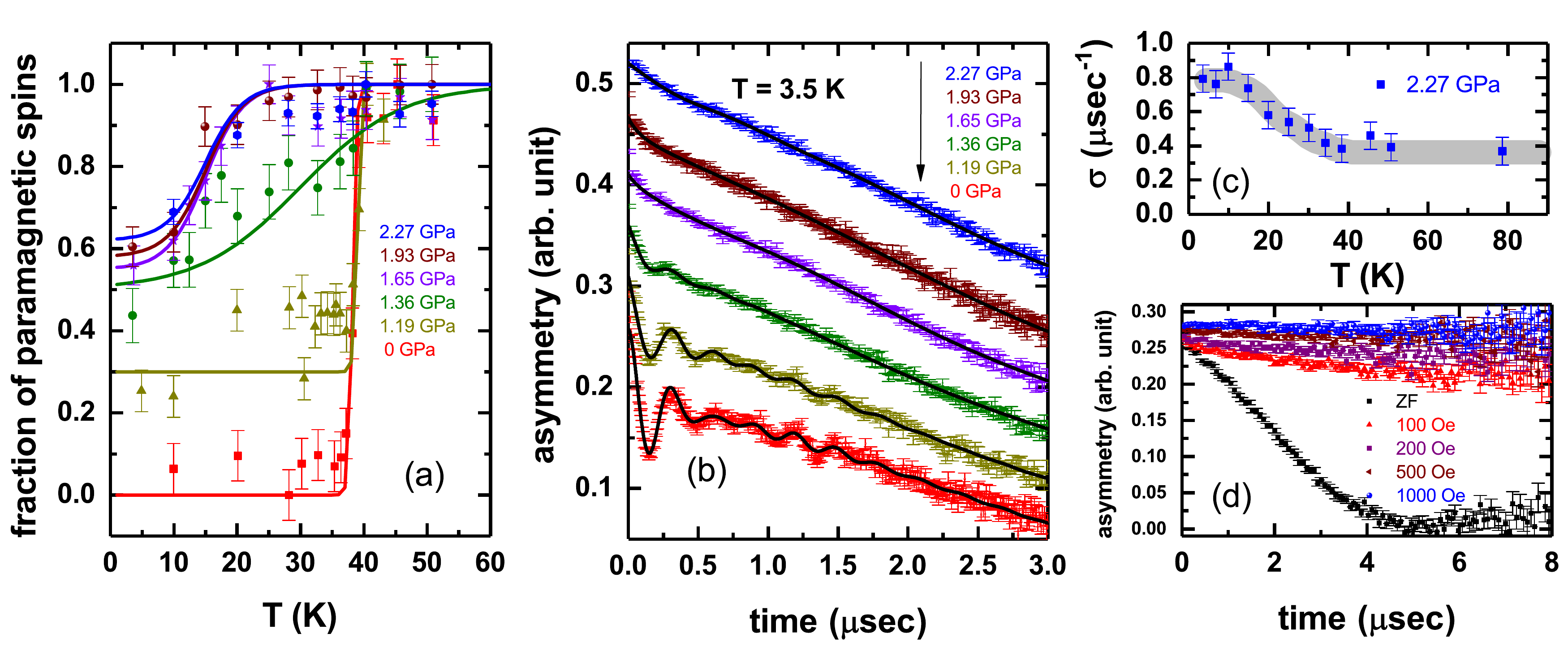}}\par} \caption{(a): Volume fraction of dynamic spins measured in the wTF experiment; the solid lines are sigmoidal fits, (b): zero-field $\mu$SR time spectra at 3.5\,K, (c): field distribution for the dynamic spins $\sigma$ as a function of temperature at 2.27\,GPa; (d): muon asymmetry in different logitudinal fields at the pressure of 2.14\,GPa at 2\,K.} \label{fig:musr}
\end{figure*}

{\cdbl\textit{Thermal expansion and magnetostriction.}} Evolution of the magnetic order under pressure was cross-checked by ambient-pressure thermal expansion measurements performed on the pressed pellet of $\beta$-Li$_2$IrO$_3$. Fig.~\ref{fig:striction}a shows a $\lambda$-like peak in the thermal expansion ($\alpha$) indicating a second-order phase transition with a non-negligible magneto-elastic coupling. The initial slope of $T_N$ is obtained from the Ehrenfest relation $dT_N/dp=V_{\rm mol} \times T_N \times \Delta \beta/\Delta C$. In our case, $V_{\rm mol}=3.354 \times 10^{-5}$~m$^3$/mol and volume expansion coefficient $\beta=3\alpha$ with $\Delta\alpha=-(0.5\pm 0.05)\times 10^{-6}$~K$^{-1}$ yields an initial pressure dependence of the transition temperature $(dT_N/dp)_{p \rightarrow 0}=(0.7\pm 0.02)$~K/GPa in agreement with the magnetization data. This confirms the positive sign of $dT_N/dp$.

Figure~\ref{fig:striction}b shows the magnetostriction coefficient, $\lambda=d[\Delta L(T)/L_0]/dB$~\cite{Knafo2009,Weickert2012} at 4\,K, as a function of magnetic field. The hump around 2.7\,T develops below $T_N$, indicating magnetoelastic coupling in the ordered state. The negative sign of $\lambda$ implies that the magnetization should increase upon compression following the Maxwell relation $\lambda\,V=-(dM/dp)_{T,B}$, where $V$ is the volume and $M$ is the magnetization. This further supports the increase in $M$ and the reduction in $H_c$ under pressure. 

The field $H_c$ marks an instability of the incommensurate state~\cite{ruiz2017,rousochatzakis2018}. The reduction in $H_c$ upon compression implies that the ambient-pressure magnetic order becomes destabilized and should eventually disappear, as we observe indeed. However, neither the magnetization data nor XMCD elucidate the nature of the high-pressure phase formed above 1.4\,GPa. The low magnetization and the absent XMCD signal could imply: i) a robust antiferromagnetic order that is not polarized by the field of several Tesla, as in $\alpha$-Li$_2$IrO$_3$ and Na$_2$IrO$_3$; ii) a dynamic spin state; iii) magnetism collapse due to, e.g., dimerization~\cite{hermann2018} or metallization. In the following, we use $\mu$SR as a sensitive local probe that distinguishes between these different scenarios and gives strong evidence for the formation of a dynamic spin state, albeit hindered by partial spin freezing.

{\cdbl\textit{$\mu$SR results.}} Muon spin relaxation ($\mu$SR) experiments were performed on polycrystalline samples. We discuss the ambient-pressure data first. At temperatures below $T_N$, $\mu$SR spectra exhibit well-defined oscillations, which indicate long-range magnetic order. Given the complex incommensurate order, a non-trivial function of the asymmetry decay can be expected. However, after trying several functions we have found that a simple sum of three cosines with the oscillation frequencies of 2.7, 3.3 and 4\,MHz reproduces the spectrum quite well~\footnote{A sum of two sites in a helical state, similar to the case of MnGe~\cite{martin2016}, reproduced the spectrum almost equally well, but introduced unnecessary complications into the analysis.},
\begin{equation}
A(t) = \dfrac{2}{3}\sum_{i=1}^3 A_i\cos (\omega_i t + \phi)\,e^{-\lambda_Tt}+\dfrac{1}{3}\,e^{-\lambda_Lt},
\end{equation}
where $\lambda_T$ and $\lambda_L$ represent, respectively, the transverse and longitudinal relaxation rates, $\omega$ is the oscillation frequency, and $\phi\simeq 0$ is the phase. The temperature dependence of the frequencies follows a phenomenological relation $\omega (T) = \omega (0)[1-(T/T_N)^{\alpha'}]^{\beta'}$ with $\alpha'\simeq 4.6$ and $\beta'\simeq 0.5$~\cite{supplement}. This $\beta'$ value indicates a mean-field type magnet, whereas the large $\alpha'$ value supports a complex magnetic state~\cite{smidman2013}.

Experiments performed with a weak transverse field (wTF) of 50\,G give access to the model-independent evaluation of the transition temperature and magnetically ordered volume fraction~\footnote{While the main results obtained using the WTF are quite robust, the determination of the volume fraction needs to be taken with caution if static fields in the sample are of the order of the applied field. Hence, the ordered fraction in the figure (color plot) in the high-pressure phase is only indicative but not absolute. It is to be noted that the background contribution has been subtracted to get 100\% magnetic volume fraction at ambient pressure.}. In the presence of the wTF, static spins do not contribute to the oscillating signal, and the asymmetry directly measures the fraction of dynamic spins in the sample. Figure~\ref{fig:musr}(a) indicates that at ambient pressure this fraction sharply drops down to zero at $T_N\simeq 38$\,K. At higher pressures, less than half of the spins become static, whereas the remaining ones are dynamic down to the lowest temperature probed in our experiment.

The crossover temperature, where part of the spins becomes static, was estimated by fitting the temperature dependence of the non-magnetic volume fraction with a sigmoidal function. We detect a slight increase in $T_N$ at the pressure of 1.19\,GPa, in agreement with the magnetization and thermal expansion data. Upon further compression, the crossover temperature decreases to about 15\,K. It no longer represents the magnetic ordering temperature, because static spins form a glassy state. This is evidenced by the zero-field data measured at 3.5\,K (Fig.~\ref{fig:musr}b). The oscillations due to the long-range-ordered state remain at the same frequencies, but reduce in magnitude upon compression and vanish above 1.37\,GPa.

We now turn our attention to the nature of the high-pressure magnetic state. The signal at high pressures is described by a sum of an oscillating function and a Gaussian relaxing function. The total asymmetry includes two contributions, one from the frozen part ($A_{\rm fr}$)~\footnote{We used an oscillatory function for the frozen spins and arrived at zero oscillation frequency that corresponds to the absence of static fields typical for a glassy state, as opposed to the non-zero frequencies characteristic of the long-range-ordered state below 1.37\,GPa.} and the other one $(1-A_{\rm fr})$ that is described by a Gaussian relaxation component $e^{-(\sigma t)^2/2}$, where $\sigma$ represents the width of the local field distribution. The $A_{\rm fr}$ has been estimated from the wTF measurements. The width of the local magnetic field is estimated to be about 10\,G at 4\,K. A longitudinal magnetic field, which is 10 times higher than that, should decouple the muon relaxation channel completely. However, even at a longitudinal magnetic field of 500\,Oe a weak relaxation survives (Fig.~\ref{fig:musr}d), which implies that correlations of unfrozen spins are dynamic in nature. The extracted temperature dependence (Fig.~\ref{fig:musr}c) shows an increase in $\sigma$ below 30\,K, indicating the onset of short-range correlations between the dynamic spins, and parallels the formation of frozen spins. Below 15\,K, both $\sigma$ and the fraction of frozen spins remain constant, indicating phase separation of $\beta$-Li$_2$IrO$_3$ into frozen spins (spin glass) and dynamic spins (spin liquid).

\begin{figure}[h]
\includegraphics{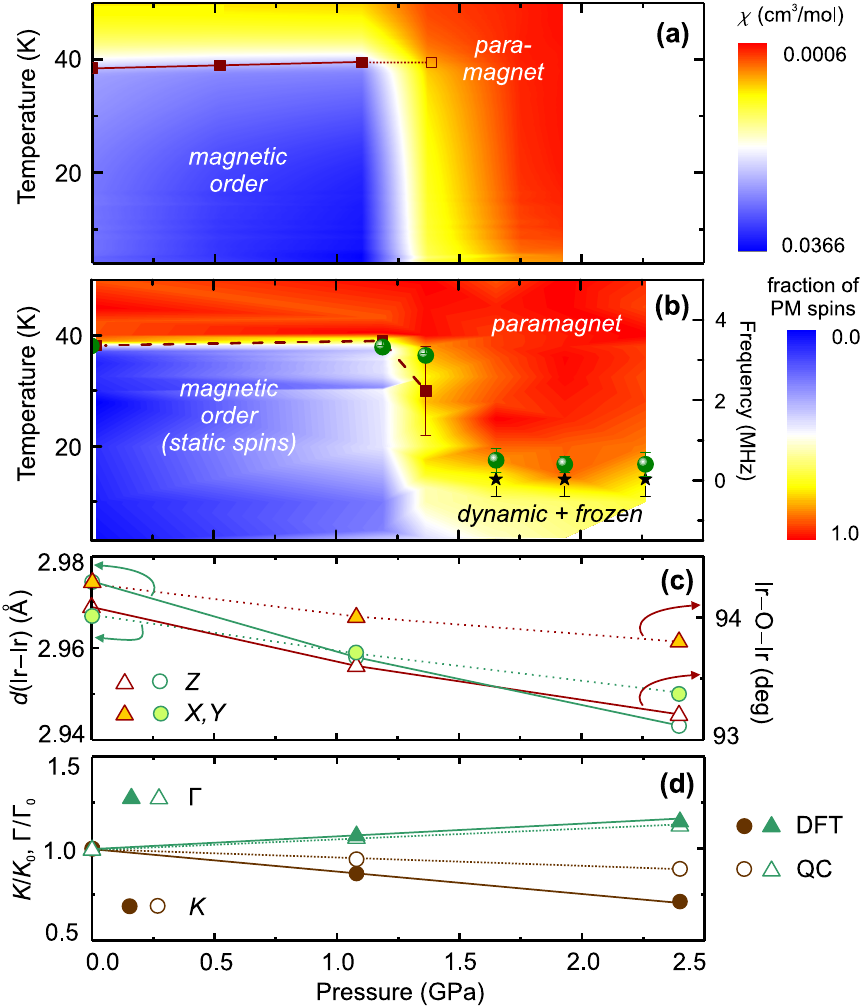} 
\caption{Phase diagram of $\beta$-Li$_2$IrO$_3$ as a function of pressure according to (a) susceptibility measurements and (b) $\mu$SR data. The circles, squares, and stars correspond, respectively, to the oscillation frequency, $T_N$, and to the temperature below which the phase-separated state occurs. Panel (c) shows the changes in the nearest-neighbor Ir--Ir distances and Ir--O--Ir angles under pressure for the $X,Y$ and $Z$-type bonds~\cite{supplement}. Panel (d) shows relative changes in the Kitaev exchange $K$ and off-diagonal anisotropy $\Gamma$ with respect to their ambient-pressure values $K_0$ and $\Gamma_0$, respectively. The $K$ and $\Gamma$ values are averaged over the $X$-, $Y$- and $Z$-type bonds. The open and field symbols are from quantum-chemical (QC) and DFT calculations, respectively. } \label{structure}
\end{figure}


{\cdbl\textit{Crystal structure and exchange couplings.}} Single-crystal x-ray diffraction (XRD) performed under pressures up to 3.45\,GPa does not reveal any drastic structural changes and excludes structural dimerization~\cite{hermann2018}, either macroscopic or local, as the possible cause for the absence of magnetic order above 1.37\,GPa. The orthorhombic symmetry of $\beta$-Li$_2$IrO$_3$ is preserved, and the Ir displacement parameters remain unchanged under pressure~\cite{supplement}. The Ir--Ir distances were extracted directly from the XRD data, whereas oxygen positions were additionally refined \textit{ab initio}~\cite{supplement}, resulting in a smooth pressure dependence. Not only the Ir--Ir distances are shortened, but also the Ir--O--Ir angles are reduced by nearly $1^{\circ}$ upon compression to 2.4\,GPa (Fig.~\ref{structure}c). 

The effect of this structural evolution was examined by electronic-structure calculations employing two complementary approaches, i) second-order perturbation theory for an effective model parametrized from density-functional (DFT) calculations~\cite{winter2016}; and ii) multireference quantum chemistry calculations for finite embedded clusters~\cite{katukuri2016}. Both methods agree on the qualitative trends for nearest-neighbor exchange couplings, namely, the absolute values of the off-diagonal exchange interaction $\Gamma$ increases, whereas the Kitaev exchange $K$ decreases (Fig.~\ref{structure}d). The Heisenberg $J$, as well as all couplings beyond nearest neighbors remain weaker than the nearest-neighbor $\Gamma$ and $K$~\cite{supplement}. 

{\cdbl\textit{Discussion.}} $\beta$-Li$_2$IrO$_3$ reveals dissimilar trends upon compression. 
The increasing $T_N$ indicates growing energies of exchange couplings, as the Ir--Ir distances shorten. In contrast, the reduction in $H_c$ points to a destabilization of the ambient-pressure magnetic order. Recent theory work~\cite{ducatman2018} considers $\beta$-Li$_2$IrO$_3$ from the perspective of two competing ordering modes. The ambient-pressure incommensurate order is due to a $Q\neq 0$ mode, which is predominant in zero field. Magnetic field applied along the $b$ direction amplifies the $Q=0$ mode and reduces the magnitude of the $Q\neq 0$ mode, eventually destroying incommensurate order above $H_c$. Our data evidence the stabilization of the $Q=0$ mode and destabilization of the $Q\neq 0$ mode also under pressure, which is concomitant with the reduction in $K$ and the increase in $\Gamma$, as our \textit{ab initio} results show (Fig.~\ref{structure}d). Around 1.4\,GPa, the $Q\neq 0$ mode is no longer active, and the incommensurate order disappears.

Two scenarios of this breakdown can be envisaged. According to Ref.~\onlinecite{ducatman2018}, the $\Gamma/K>1$ region should be characterized by another type of magnetic order, which may appear in the narrow pressure range around 1.4\,GPa before the spin-liquid phase of the large-$\Gamma$ limit~\cite{rousochatzakis2017} sets in. However, our data are also consistent with a direct, first-order transformation between the incommensurate order and spin liquid, similar to the pressure-induced breakdown of magnetic order in itinerant magnets, where phase separation is typically observed, with ordered and disordered phases coexisting in a broad pressure range~\cite{uemura2007}. Indeed, at low temperatures we observe a fraction of disordered spins already at 1.19\,GPa, as well as the coexistence of ordered and disordered states at 1.36\,GPa, thus confirming the first-order nature of the transition (Fig.~\ref{fig:musr}a). The reduced magnetization at 1.37\,GPa (Fig.~\ref{fig:chi}) would be then due to the coexistence of the ordered phase and spin liquid. 

Interestingly, the ground state of $\beta$-Li$_2$IrO$_3$ well above 1.4\,GPa is also phase-separated, but this time it represents a mixture of two disordered states, spin liquid and spin glass. Similar features have been seen in powder samples of the kagome mineral vesigneite~\cite{colman2011}, although single crystals of the same mineral show clear signatures of a magnetic transition~\cite{ishikawa2017}, thus hinting at the structural disorder as the origin of both dynamic spin state and partial freezing therein. 

$\beta$-Li$_2$IrO$_3$ is clearly different, because it does show robust magnetic order at ambient pressure and, according to XRD data~\cite{supplement}, lacks any visible structural defects, either native or pressure-induced. Therefore, we are led to conclude that dynamic spins in pressurized $\beta$-Li$_2$IrO$_3$ represent a spin-liquid state, but this liquid is highly fragile. Strong tendency toward freezing is more likely to occur in a classical spin liquid, which is indeed anticipated in the large-$\Gamma$ limit that our system approaches. With $\Gamma<0$, exchange terms beyond $\Gamma$ should cause order by disorder, but its energy scale is as low as $\Gamma/64\leq 3$\,K, going beyond the lower limit of our data.

{\cdbl\textit{Conclusions.}} Incommensurate magnetic order in $\beta$-Li$_2$IrO$_3$ is destabilized under pressure and vanishes upon the first-order transition around 1.4\,GPa, giving way to the coexisting dynamic and static spins in a partially frozen spin liquid. A plausible explanation of this effect is the formation of a classical spin liquid prone to spin freezing. Such a state is indeed expected in the large-$\Gamma$ limit that $\beta$-Li$_2$IrO$_3$ tends to approach. Our results do not support the pressure-induced formation of a quantum spin liquid, and instead put pressurized $\beta$-Li$_2$IrO$_3$ forward as a suitable platform for studying classical spin liquid in the large-$\Gamma$ limit of the extended Kitaev model, an interesting and hitherto sunexplored field. The natural next step in this endeavor would be nuclear magnetic resonance and electron spin resonance measurements probing spin dynamics in pressurized $\beta$-Li$_2$IrO$_3$ on different time scales.

\acknowledgments
AAT is indebted to Ioannis Rousochatzakis for insightful conversations and sharing his unpublished results. We also acknowledge fruitful discussions with Steve Winter and Radu Coldea, and the provision of the $\mu$SR beamtime by PSI and ISIS. RSM would like to thank Dr. Naoyuki Tateiwa and Dr. Yoshifumi Tokiwa for their suggestions on optimizing parameters of the pressure cell. The work in Augsburg was supported by DFG under TRR80 (FF, AAT, PG) and JE-748/1 (AJE), and by the Federal Ministry for Education and Research through the Sofja Kovalevskaya Award of Alexander von Humboldt Foundation (MM, TD, AAT). The work of GS is supported by the Swiss National Science foundation grants 200021\_149486 and 200021\_175935.

%

\clearpage\newpage
\begin{widetext}
\begin{center}
\large\textbf{\textit{Supplemental Material}\smallskip \\ Breakdown of magnetic order in the pressurized Kitaev iridate $\beta$-Li$_2$IrO$_3$}
\end{center}
\end{widetext}

\renewcommand{\thefigure}{S\arabic{figure}}
\renewcommand{\thetable}{S\arabic{table}}
\setcounter{figure}{0}

\section{Sample preparation}
Polycrystalline samples of $\beta$-Li$_2$IrO$_3$ were synthesized from stoichiometric mixtures of Li$_2$CO$_3$ and IrO$_2$ in air at 1050\,$^{\circ}$C with several intermediate re-grindings. X-ray diffraction (XRD) data showed no impurity phases except for about 1\,wt.\% of $\alpha$-Li$_2$IrO$_3$ that does not affect any of the results presented in the manuscript.

Single crystals of $\beta$-Li$_2$IrO$_3$ were grown from separated educts~\cite{freund2016} at 1020\,$^{\circ}$C using Li and Ir metals as starting materials. 

\section{X-ray diffraction}
Powder XRD data were collected using the MiniFlex (Rigaku) and Empyrean (PANalytical) diffractometers with the CuK$_{\alpha}$ radiation. JANA2006 software~\cite{jana2006} was used for the Rietveld refinement. Unlike $\alpha$-Li$_2$IrO$_3$, which is prone to twinning and stacking faults~\cite{freund2016}, \mbox{$\beta$-Li$_2$IrO$_3$} shows high degree of crystallinity and the symmetric peak shape. The powder profile could be fitted by a Lorentzian function without any additional corrections for strain broadening. Moreover, the peak width defined by the two Lorentzian parameters remained essentially unchanged after the powder sample was subjected to pressure treatment in the course of the $\mu$SR experiment (Table~\ref{tab:refinement}). This indicates a low amount of structural defects in the material and the absence of pressure-induced defects.

Single-crystal XRD data were collected at the ID27 beamline of the ESRF, Grenoble, France (Perkin Elmer XRD1621 flat panel detector, sample-to-detector distance 418.8\,mm, $\lambda=0.3738$\,\r A, x-ray spot size $2.6\times 2.6$\,$\mu$m$^2$). XRD images were collected during continuous rotation of the diamond anvil cell (DAC), typically from $-20$ to $+20^{\circ}$ on $\omega$; while data collection experiments were performed by a narrow $0.5^{\circ}$ scanning of the omega range from $-30$ to $+30^{\circ}$. 

The DAC equipped with 350\,$\mu$m Boehler-Almax diamonds was used for pressure generation. Two single crystals of $\beta$-Li$_2$IrO$_3$ of about $15\times 15\times 10$\,$\mu$m$^3$ size, together with a small ruby chip (for pressure determination), were loaded into a hole of a pre-indented rhenium gasket. Helium was used as a pressure-transmitting medium. Only one of the two crystals had the quality sufficient for structure solution and refinement.

Integrations of the reflection intensities were performed using the CrysAlisPro software~\cite{crysalispro}. A single crystal of an orthoenstatite [(Mg$_{1.93}$,Fe$_{0.06}$)(Si$_{1.93}$,Al$_{0.06}$)O$_6$, $Pbca$, $a=8.8117(2)$, $b=5.18320(10)$, $c=18.2391(3)$\,\r A], was used to calibrate the instrument model of CrysAlisPro software (sample-to-detector distance, the detector's origin, offsets of the goniometer angles and rotation of the X-ray beam and the detector around the instrument axis).

\begin{figure}
\includegraphics[width=8.7cm]{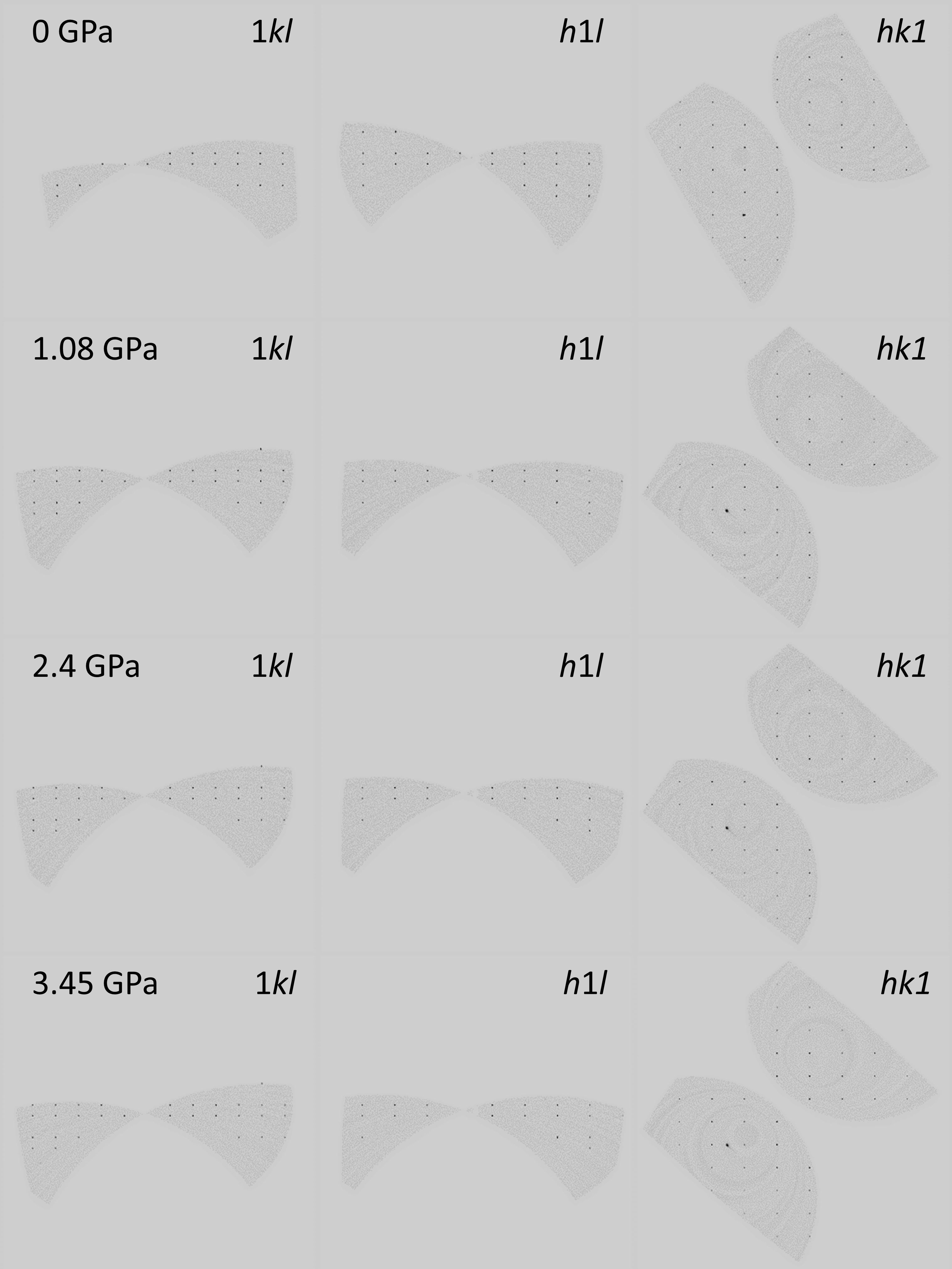}
\caption{\label{fig:xrd}
Reciprocal space images of the $\beta$-Li$_2$IrO$_3$ crystal under pressure. 
}
\end{figure}

The structures were solved with SHELXT~\cite{shelxt} and refined against $F^2$ on all data by the full-matrix least squares method with SHELXL~\cite{shelxl}. During the compression, no significant change in the crystal quality ($R_{\rm int}$ and sample mosaicity) was observed. The absolute values of the sample mosaicity representing average peak widths remain almost unchanged (Table~\ref{tab:xrd2}) confirming that no pressure-induced defects occur in $\beta$-Li$_2$IrO$_3$. Note that the $e_3$ parameter describes the reflection width in the scanning ($\omega$) direction, therefore it is always larger than the scan width ($0.5^{\circ}$). Reciprocal space images (Figure~\ref{fig:xrd}) further confirm the unchanged crystal quality under pressure. 

\begin{table*}
\caption{\label{tab:xrd2}
Integration quality ($R_{\rm int}$) and crystal mosaicity ($e_1,e_2,e_3$) in single-crystal XRD measurements.
}
\begin{ruledtabular}
\begin{tabular}{ccccc}
 & 0\,GPa & 1.08\,GPa & 2.40\,GPa & 3.45\,GPa \\
$R_{\rm int}$ (\%) & 4.80 & 6.46 & 6.19 & 5.12 \\
$e_1/e_2/e_3$ & 0.11/0.11/0.58 & 0.11/0.11/0.52 & 0.11/0.11/0.50 & 0.12/0.11/0.57
\end{tabular}
\end{ruledtabular}
\end{table*}
\begin{table}
\caption{\label{tab:refinement}
Rietveld refinement results for $\beta$-Li$_2$IrO$_3$ samples before and after pressure treatment in the course of the $\mu$SR experiment. The lattice parameters $a$, $b$, and $c$, as well as the Lorentzian profile parameters LX and LY are listed. The error bars are from the least-squares refinement against all data points of the XRD profile and thus smaller than the actual error bars.
}
\begin{ruledtabular}
\begin{tabular}{cccccc}
 Sample & $a$        & $b$        & $c$        & LX      & LY     \\\hline
 before & 5.90444(4) & 8.44958(7) & 17.8117(2) & 5.12(7) & 4.3(2) \\
 after  & 5.90470(3) & 8.45012(5) & 17.8128(1) & 4.88(6) & 5.1(1) \\
\end{tabular}
\end{ruledtabular}
\end{table}

\section{Magnetization measurements}
The bulk magnetization measurements under hydrostatic pressure~\cite{tateiwa2011,tateiwa2013} were performed on a polycrystalline sample in a CuBe pressure cell placed inside a commercial SQUID magnetometer from Quantum Design. All measurements were performed upon compression, whereas no pressure control during decompression was possible. The highest feasible pressure was about 2.0\,GPa. Daphne 7373 oil was used as pressure transmitting medium. One small piece of lead ($\sim0.1$\,mg) is placed together with the sample inside the pressure cell and another piece ($\sim0.1$\,mg) is placed outside the pressure cell. Under pressure the superconducting transition temperatures of the inner lead sample decreases. The difference between the superconducting transition temperatures of the two lead samples determines the pressure inside the cell. 
The gasket of the pressure cell contained the sample of mass $\sim1$\,mg and a lead piece of mass $\sim0.1$\,mg. The empty cell background data has been subtracted~\cite{tateiwa2011} by using automatic background subtraction (ABS) procedure mentioned in Ref.~\onlinecite{MPMS}. Measurements of lead and sample were performed with the fields of 2\,mT and 1\,T respectively. Measurements under pressure have been reproduced several times.

Field-cooled and zero-field-cooled scans were performed at ambient pressure and showed no dependence on the cooling regime. Above 1.4\,GPa, the signal of the sample was at the sensitivity limit of our measurement setup even in the field of 1\,T, so we can neither confirm nor exclude the dependence of the magnetization on the cooling history, as expected in the glassy phase pinpointed by $\mu$SR below 15\,K.

\section{Specific heat and thermal expansion}

Specific heat measurements were carried out on a polycrystalline sample in the Quantum Design PPMS with thermal relaxation method. Thermal expansion was measured by high-resolution capacitive dilatometry, enabling the detection of length changes $\Delta L(T)$ smaller than $0.05$\,\AA~over a sample with the length $L_0$ of several\,mm~\mbox{\cite{barron1980,manna2012,kuechler2012}}. We utilized the dilatometer of Ref.~\onlinecite{kuechler2012} in the multi-function probe of the PPMS. The linear thermal expansion coefficient $\alpha=d[\Delta L(T)/L_0]/dT$ was determined from the differential length change over temperature intervals of 0.5\,K. Measurements were done on a pressed pellet of 2.1\,mm length. Pellets were pressed inside the glove box in order to avoid air trapping inside the pellet. Two different pellets from different batches have been measured in order to check the reproducibility. Thermal expansion data were taken upon warming with a temperature sweep rate of $+0.3$\,K/min. Isothermal field sweeps, i.e., magnetostriction measurements were performed up to 14\,T with a field sweep rate of $+120$\,mT/min.

\section{$\mu$SR experiments}

Ambient-pressure $\mu$SR experiments were carried out at the HIFI beam of ISIS, UK and Dolly spectrometer of PSI, Switzerland. Pressure experiments were performed at the GPD spectrometer of PSI, Switzerland. The $\mu$SR time spectra were analyzed using the MUSRFIT software package. A 2\,g polycrystalline sample was used.

To generate high pressure, a double-wall piston-cylinder type cell manufactured from MP35 alloy was used~\cite{khasanov2016}. This allowed for a significant sample volume, high enough pressure, and temperature-independent background in the range studied. The momentum of incoming muons was chosen to optimize the stopping of the muons within the sample area. In order to transmit and distribute the pressure, Daphne7373 oil was used. The pressure was applied at room temperature. It was additionally measured at low temperatures by monitoring the pressure-induced shift of the superconducting transition temperature of indium.

The total $\mu$SR signal presented here consists of two contributions
\begin {equation}
A(t) = A_{\rm PC}P_{\rm PC}(t) + A_SP_S(t)
\end {equation}
where $A_{\rm PC}(t)$ and $A_S(t)$ represent the asymmetry of the signal coming from the pressure cell and the sample itself, and $P_{\rm PC}$ and $P_S$ corresponds to the function, which evolves with time $t$. The contribution of the signal from the pressure cell is around 50\% and has been kept constant as a function of temperature and pressure. Kubo-Toyabe depolarization function has been used as $P_{\rm PC}$. The extracted parameters are same as expected for an empty pressure cell~\cite{khasanov2016}. 

\begin{figure}
{\centering {\includegraphics[width=0.53\textwidth]{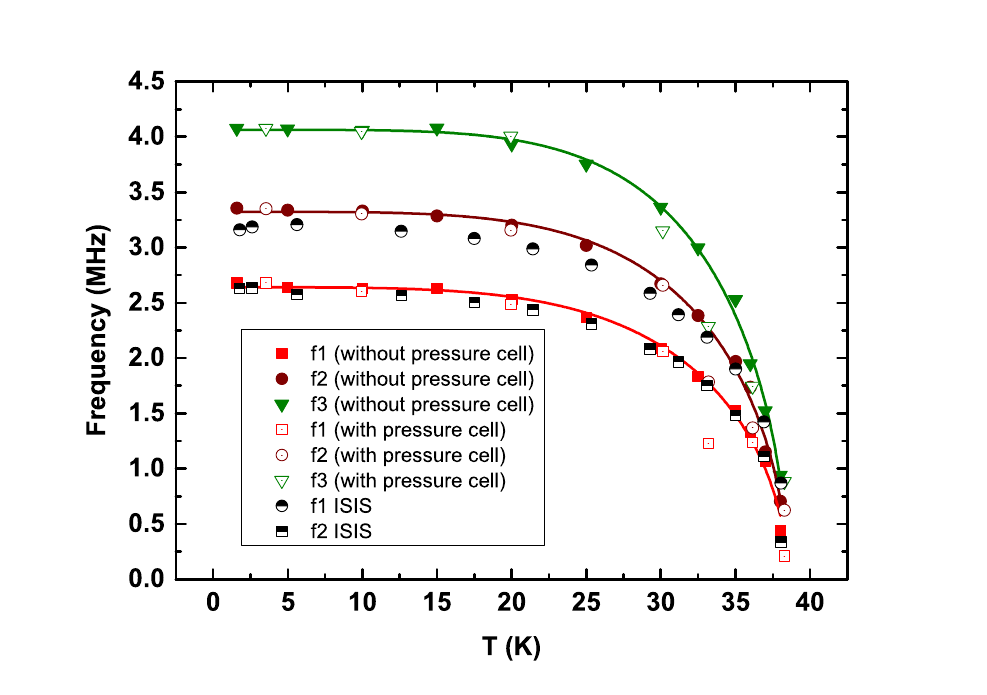}}\par} \caption{Temperature dependence of the three oscillating frequencies estimated from the measurement at ISIS and also from the measurement at PSI with pressure cell and without pressure cell, all performed at ambient pressure. The solid lines are described in the main text.} \label{fig:oscillations}
\end{figure}


At first, we performed ambient pressure experiments with the pressure cell. We used the function described in the main text as $P_S$, and confirmed that both temperature dependence and magnitudes of the three oscillation frequencies are same in nature as in the measurements without the pressure cell (Fig.~\ref{fig:oscillations}). It has to be mentioned that three oscillation frequencies have been resolved from the PSI data, whereas only two were visible in the ISIS data, but they are quite close to the PSI ones. The third, highest frequency was overlooked at ISIS due to the fact that PSI can accumulate muons at much lower time compared to that at ISIS.

\begin{figure}
{\centering {\includegraphics[width=0.4\textwidth]{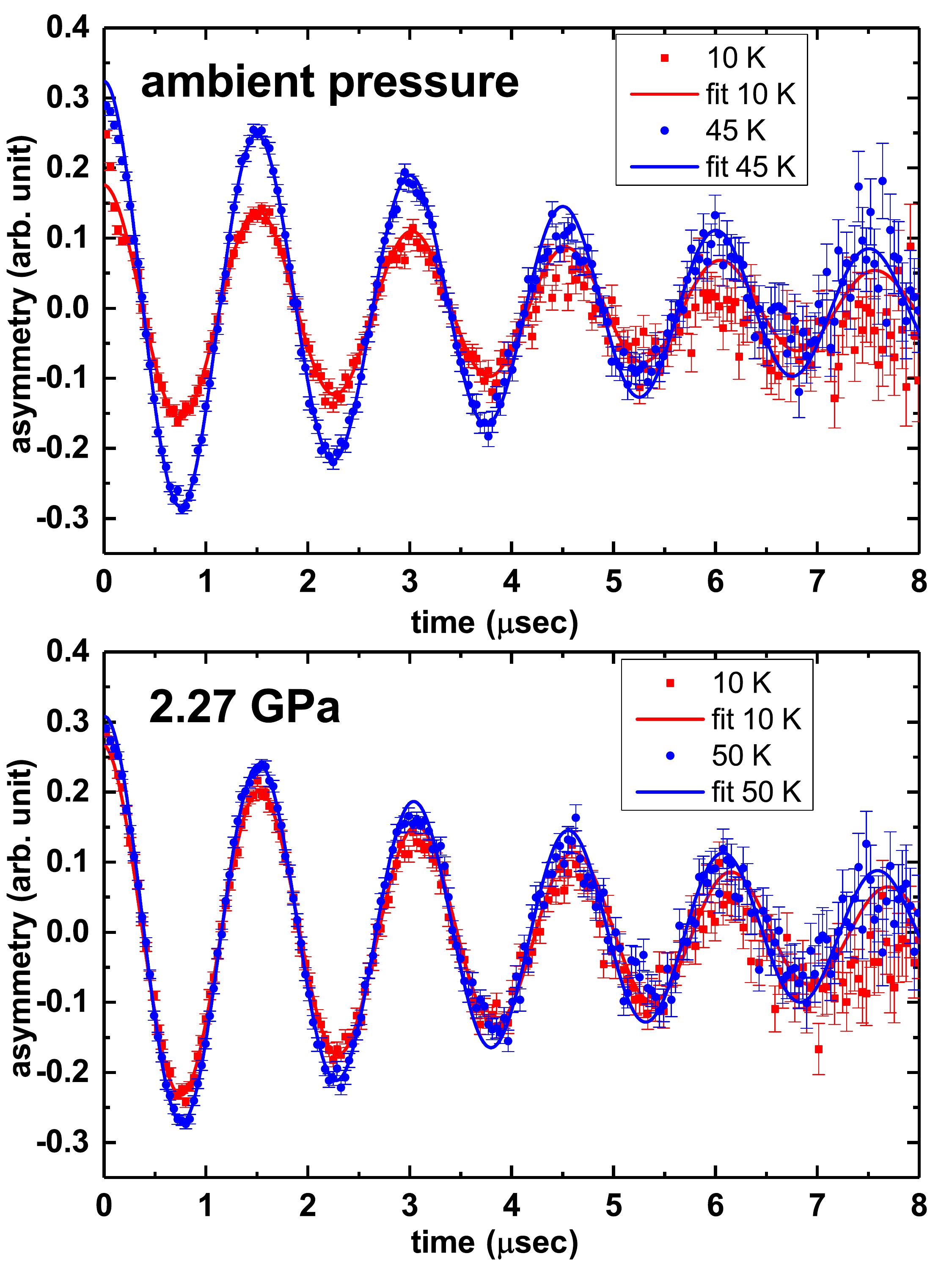}}\par} \caption{$\mu$SR time spectra in the presence of weak transverse magnetic field of 50 Gauss at ambient pressure and highest pressure, and at temperatures below and above the ordering temperatures.} \label{fig:tf}
\end{figure}


The upper panel of Figure~\ref{fig:tf} displays the weak transverse field (WTF) measurements at ambient pressure at 45\,K (above the ordering temperature) and 10\,K (below the ordering temperature), whereas the lower panel shows the same data for the pressure of 2.27\,GPa. To deconvolute the $\mu$SR time spectra, we have used the following equation
\begin{equation}
A(t) = A_0 \cos(\omega t)\, e^{-\lambda t}
\end{equation}
where $\omega$ corresponds to the field of 50\,G. The temperature dependence of $A(t)$ gives the non-magnetic volume fraction. From Figure~\ref{fig:tf} it can be clearly seen that the volume fraction of static spins at 10\,K decreased for the applied pressure of 2.27\,GPa compared to the ambient pressure. This is compatible with the increased fraction of paramagnetic spins in Fig.~3a of the manuscript.  

\section{Crystal structure under pressure}
Precise structure determination for $\beta$-Li$_2$IrO$_3$ is hindered by the large difference in the scattering powers of Ir and light elements (Li and O). Ambient-pressure crystal structures reported in Refs.~\onlinecite{biffin2014,takayama2015} show small, but not insignificant differences in the lattice parameters and oxygen positions, which in turn influence Ir--O distances and Ir--O--Ir angles underlying magnetic exchange. For example, Ref.~\onlinecite{takayama2015} reports nearly ideal IrO$_6$ octahedra with the Ir--O distances of 2.025(3), 2.024(3), and 2.027(3)\,\r A, whereas the crystal structure of Ref.~\onlinecite{biffin2014} displays a somewhat asymmetric oxygen environment with the Ir--O distances of 2.01(3), 2.04(2), and 2.07(4)\,\r A. Moreover, the Ir--O--Ir angles for the $X/Y$- and $Z$-type Kitaev bonds are, respectively, $94.4(1)^{\circ}$ and $94.7(2)^{\circ}$ in Ref.~\onlinecite{takayama2015}, compared to $92.5(4)^{\circ}$ and $95.2(7)^{\circ}$ in Ref.~\onlinecite{biffin2014}

While the difference in the lattice parameters can be traced back to different temperatures of the experiment (100\,K~\cite{biffin2014} vs. 296\,K~\cite{takayama2015}), atomic parameters are unlikely to change significantly upon cooling. Their discrepancies are a result of the lower precision of the oxygen positions in Ref.~\onlinecite{biffin2014}, which is also seen from the much higher error bars. This problem is rooted in the lower number of independent reflections used in the refinement, 298 vs. 1248.

Pressure cell reduces the accessible part of the reciprocal space and thus the number of independent reflections to about 105 in our data. While this would be enough to refine 7 structural parameters (positions of Li, Ir, and O) and 3 atomic displacement parameters (Ir, O1, and O2), we expect only a moderate accuracy for the positions of light atoms. Indeed, the refinement of the XRD data leads to realistic Ir--O distances of $2.0-2.1$\,\r A, but the Ir--O--Ir angles show a large scatter and no systematic change under pressure, similar to Ref.~\onlinecite{veiga2017}. To circumvent this problem, we undertook a combined approach, with the lattice parameters and Ir positions determined by XRD, while Li and O positions were refined \textit{ab initio}.

\begin{table}
\caption{\label{tab:structures}
Crystallographic parameters of $\beta$-Li$_2$IrO$_3$ under pressure listed for the space group $Fddd$ (setting 2). The $z$-coordinate and atomic displacement parameter $U_{\rm iso}$ (in\,\r A$^2$) of Ir are determined from the refinement of single-crystal XRD data. The oxygen and Li positions are further refined \textit{ab initio}, as explained in the text. The last two lines list \mbox{Ir--Ir} distances (in\,\r A) and Ir--O--Ir bridging angles (in~deg) for the \mbox{$X$,$Y$-/$Z$-}type bonds of the hyperhoneycomb lattice (see also Fig.~\ref{fig:structure}).
}
\begin{ruledtabular}
\begin{tabular}{ccccc}
 & 0\,GPa & 1.08\,GPa & 2.4\,GPa & 3.45\,GPa \\
$a$ (\r A) & 5.9004(3) & 5.8816(3) & 5.8614(3) & 5.8475(4) \\
$b$ (\r A) & 8.4457(5) & 8.4054(5) & 8.3590(4) & 8.3147(5) \\
$c$ (\r A) & 17.795(14) & 17.736(17) & 17.687(15) & 17.618(19) \\
$z$(Ir) & 0.7086(2) & 0.7084(2) & 0.7082(3) & 0.7086(3) \\\smallskip
$U_{\rm iso}$(Ir) & 0.016(3) & 0.010(3) & 0.010(4) & 0.009(3) \\
$x$(O1) & 0.8596 & 0.8612 & 0.8642 & 0.8625 \\
$x$(O2) & 0.6316 & 0.6314 & 0.6312 & 0.6303 \\
$y$(O2) & 0.3648 & 0.3657 & 0.3667 & 0.3676 \\
$z$(O2) & 0.0384 & 0.0386 & 0.0387 & 0.0390 \\
$z$(Li1) & 0.0454 & 0.0454 & 0.0453 & 0.0457 \\\smallskip
$z$(Li2) & 0.8718 & 0.8783 & 0.8781 & 0.8782 \\
$d_{\rm Ir-Ir}$ & 2.967/2.975 & 2.959/2.958 & 2.950/2.943 & 2.930/2.946 \\
$\varphi_{\rm Ir-O-Ir}$ & 94.3/94.1 & 94.0/93.6 & 93.8/93.2 & 93.2/93.4 \\
\end{tabular}
\end{ruledtabular}
\end{table}

VASP code~\cite{vasp1,vasp2} was used for crystal structure optimization. Details of the relaxed structures strongly depend on the underlying approximation. We tested several exchange-correlation potentials and different settings for the spin-orbit coupling and correlation effects. Similar to Ref.~\onlinecite{hermann2018}, calculations without the spin-orbit coupling and Hubbard $U_d$ resulted in structural dimerization. Experimental ambient-pressure crystal structure of Ref.~\onlinecite{takayama2015} is well reproduced only on the DFT+$U$+SO level, whereas the choice of the exchange-correlation potential and the change in the $U_d$ value had only a minor effect. The best agreement was found for $U_d=3$\,eV, $J_d=0.5$\,eV, and the PBEsol exchange-correlation potential~\cite{perdew2008}. Same methodology was then used for relaxing Li and O positions under pressure, whereas Ir atoms were placed into their experimental positions and fixed.

The resulting structural parameters are summarized in Table~\ref{tab:structures}. Two effects are worth noting. First, by combining XRD determination of the Ir position with the \textit{ab initio} refinement of Li and O coordinates, we obtain a rather monotonic evolution of the Ir--O--Ir angles, which then allows to track changes in the exchange couplings. The structural changes are well in line with earlier predictions based on DFT~\cite{kim2016}. Second, the atomic displacement parameter of Ir remains nearly constant under pressure, thus excluding the formation of local Ir--Ir dimers up to at least 3.45\,GPa. This proves that the spin-liquid state pinpointed in our $\mu$SR experiment occurs on the undistorted hyperhoneycomb lattice of $\beta$-Li$_2$IrO$_3$.

\section{Exchange couplings}
All exchange parameters are given in the global coordinate frame defined as
\begin{equation}
 \Xv=(\ev_a+\ev_c)/\sqrt 2,\,\,\, \Yv=(\ev_c-\ev_a)/\sqrt 2,\,\,\, \Zv=-\ev_b,
\end{equation}
where $\ev_a,\ev_b$, and $\ev_c$ are unitary vectors along the $a$, $b$, and $c$ crystallographic directions, respectively. 

The presence of Kitaev interactions discriminates all nearest-neighbor bonds into the $X$-, $Y$-, and $Z$-types with the symmetric part of the exchange written as follows~\cite{winter2016},
\begin{widetext}
\begin{equation*}
 \mathbb J_X\!=\!\left(\begin{array}{ccc} J_{XY}+K_{XY} & \Gamma_{XY}'+\zeta & \Gamma_{XY}'-\zeta \\ \Gamma_{XY}'+\zeta & J_{XY}+\xi & \Gamma_{XY} \\ \Gamma_{XY}'-\zeta & \Gamma_{XY} & J_{XY}-\xi \end{array}\right),\,\, \mathbb J_Y\!=\!\left(\begin{array}{ccc} J_{XY}+\xi & \Gamma_{XY}'+\zeta & \Gamma_{XY} \\ \Gamma_{XY}'+\zeta & J_{XY}+K_{XY} & \Gamma_{XY}'-\zeta \\ \Gamma_{XY} & \Gamma_{XY}'-\zeta & J_{XY}-\xi \end{array}\right),\,\, \mathbb J_Z\!=\!\left(\begin{array}{ccc} J_Z & \Gamma_Z & 0 \\ \Gamma_Z & J_Z & 0 \\ 0 & 0 & J_Z+K_Z \end{array}\right).
\end{equation*}
\end{widetext}

\begin{figure}
\includegraphics{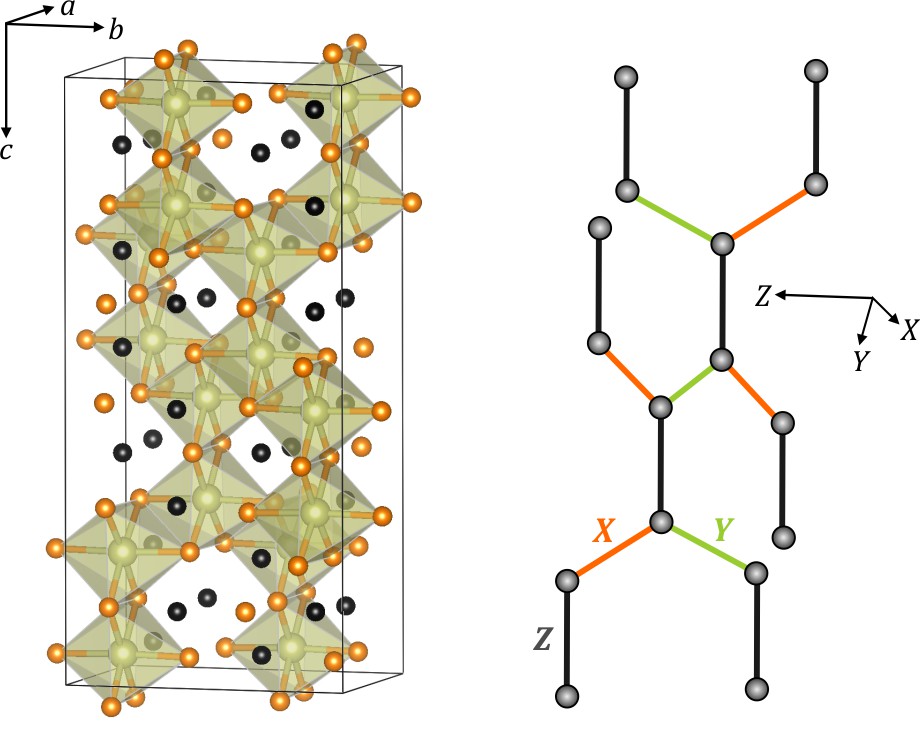}
\caption{\label{fig:structure}
Crystal structure of $\beta$-Li$_2$IrO$_3$ (left) and the hyperhoneycomb spin lattice (right). The crystallographic coordinate frame $abc$ and spin coordinate frame $XYZ$ are shown.
}
\end{figure}

The $X$- and $Y$-bonds are related by symmetry and thus feature same values of the exchange parameters, although the signs of the off-diagonal terms change from one bond to another following symmetry transformations of the $\beta$-Li$_2$IrO$_3$ structure. In particular, the $\Gamma_{XY}$-term changes sign, see Ref.~\onlinecite{ducatman2018} for further details. The $X$- and $Y$- bonds adopt the $C_i$ symmetry that forbids antisymmetric exchange.

The $Z$-bonds adopt the higher symmetry $D_2$ that sets to zero all off-diagonal terms other than $\Gamma$, and renders the sign of $\Gamma_Z$ constant throughout the lattice. Therefore, the sign of $\Gamma$ can be defined unambiguously as the sign of $\Gamma_Z$ within the given coordinate frame, and it is this sign that defines the $\Gamma<0$ or $\Gamma>0$ regimes of the $J-K-\Gamma$ model on the hyperhoneycomb lattice~\cite{lee2015,rousochatzakis2017}. Our choice of $\Xv$, $\Yv$, and $\Zv$ follows that of Refs.~\onlinecite{lee2015,rousochatzakis2017,ducatman2018}, thus facilitating a direct comparison to theory. The absence of inversion symmetry on the $Z$-bond allows an antisymmetric Dzyaloshinsky-Moriya interaction of the form $(D,D,0)$.

\subsection{DFT}
Full exchange tensors were calculated using atomic positions from Table~\ref{tab:structures} within the second-order perturbation theory~\cite{winter2016} in electronic correlations ($U_{\rm eff}=1.7$\,eV, $J_H=0.3$\,eV) and spin-orbit coupling ($\lambda=0.4$\,eV)~\footnote{Note that $U_{\rm eff}$ and $J_H$ are smaller than, respectively, $U_d$ and $J_d$ of DFT+$U$+SO, because $U_{\rm eff}$ and $J_H$ correspond to the mixed Ir--O states in the valence band, whereas $U_d$ and $J_d$ are applied to the Ir $5d$ states only.}. Hopping parameters within the $t_{2g}$ manifold were obtained in the FPLO code~\cite{fplo} on the scalar-relativistic level of local density approximation (LDA)~\cite{pw92} by constructing Wannier functions via the internal procedure of FPLO~\cite{wannier}.

\begin{table}
\caption{\label{tab:dft}
Nearest-neighbor exchange parameters (in\,meV) from second-order perturbation theory (DFT). 
}
\begin{ruledtabular}
\begin{tabular}{ccccc}
 Pressure (GPa) & $K_Z$ & $J_Z$ & $\Gamma_Z$ & $D$ \\
 0   & $-10.52$ & $-5.38$ & $-13.63$ & 0.56 \\
1.08 & $-7.74$ & $-6.60$ & $-15.28$ & 0.47 \\
2.40 & $-5.52$ & $-7.50$ & $-16.71$ & 0.40 \\
3.45 & $-5.91$ & $-7.24$ & $-15.93$ & 0.37 \\
\end{tabular}
\end{ruledtabular}
\vspace{0.1cm}

\begin{ruledtabular}
\begin{tabular}{ccccccc}
 Pressure (GPa) & $K_{XY}$ & $J_{XY}$ & $\Gamma_{XY}$ & $\Gamma_{XY}'$ & $\xi$ & $\zeta$ \\
 0   & $-12.10$ & $-4.76$ & $-13.53$ & 0.32 & $-0.10$ & 0.69 \\
1.08 & $-10.14$ & $-5.49$ & $-14.31$ & 0.21 & $-0.13$ & 0.66 \\
2.40 & $-8.80$ & $-6.11$ & $-15.06$ & 0.09 & $-0.10$ & 0.59 \\
3.45 & $-5.70$ & $-7.50$ & $-17.12$ & 0.31 & $-0.10$ & 0.51 \\
\end{tabular}
\end{ruledtabular}
\end{table}

Table~\ref{tab:dft} lists all nearest-neighbor interactions in \mbox{$\beta$-Li$_2$IrO$_3$}. Additionally, we calculated the couplings between second and third neighbors, which are all below 0.5\,meV except for $J_3$, which shows values as high as 2.66\,meV at ambient pressure and increases to 3.39\,meV at 3.45\,GPa. Although non-negligible, $J_3$ is still weaker than nearest-neighbor $\Gamma$ and $K$, which justifies the neglect of these coupling to a first approximation. 

\subsection{Quantum chemistry}
The material model was based on embedded clusters with two edge-sharing octahedra as central region. The four nearest-neighbor octahedra were also explicitly included in the calculations in order to describe the finite charge distribution in the immediate neighborhood, while the solid-state surroundings were modeled by an array of point charges fitted to reproduce the ionic Madelung potential in the cluster region. Energy-consistent relativistic pseudopotentials along with quadruple-zeta basis functions were used for the Ir\,\cite{Figgen09} ions of the central unit. All-electron basis sets of quintuple-zeta quality were employed for the bridging O\,\cite{Dunning89} ligands while all-electron basis sets of triple-zeta quality were used for the remaining O sites\,\cite{Dunning89} present in the two-octahedra central region. Ir$^{4+}$ sites belonging to the octahedra adjacent to the reference unit were described as closed-shell Pt$^{4+}$ ions, using relativistic pseudopotentials and valence triple-zeta basis functions\,\cite{Figgen09}. Ligands of these adjacent octahedra that are not shared with the central reference unit were modeled with minimal all-electron atomic-natural-orbital basis sets \cite{Pierloot95}. All calculations were performed using the quantum chemistry package {\sc molpro}\,\cite{Molpro12}.

Results of spin-orbit MRCI calculations for the nearest-neighbor effective couplings are listed in Table\,\ref{QC_table}. Details of the mapping procedure are described in, e.\,g., Ref.\,\cite{yadav16}. For the $X$- and $Y$-bonds we neglect small lattice distortions that reduce the point-group symmetry from $C_{2h}$ to $C_i$. This translates to setting $\xi$ and $\Gamma_{XY}'$ to zero, an approximation that finds support in the fact that $\xi$ and $\Gamma_{XY}'$ are the smallest parameters in the DFT-based derivation (Table~\ref{tab:dft}). 
 
\begin{table}[t]
\caption{\label{tab:qchem}
Nearest-neighbor exchange parameters (in\,meV) from quantum chemistry calculations, see text for details.}
\label{QC_table}
\begin{ruledtabular}
\begin{tabular}{ccccc}
Pressure\,(GPa) & $K_Z$  & $J_Z$ & $D$ & $\Gamma_Z$   \\
      $0$  & $-12.75$ & $-0.18$ & $0.64$ & $-2.88$ \\
   $1.08$  & $-11.80$ & $-0.53$ & $0.75$ & $-3.20$ \\
   $2.40$  & $-11.24$ & $-0.79$ & $0.81$ & $-3.56$ \\
   $3.45$  & $-11.51$ & $-0.76$ & $0.77$ & $-3.43$ \\
\end{tabular}
\end{ruledtabular}
 
\vspace{0.1cm}
\begin{ruledtabular}
\begin{tabular}{ccccc}
 Pressure\,(GPa) & $K_{XY}$ & $J_{XY}$ & $\Gamma_{XY}$ & $\zeta$ \\
   $0$  & $-12.62$ & $-0.40$ & $-3.90$ & $-0.38$ \\
 $1.08$ & $-12.01$  & $-0.76$ & $-4.11$ & $-0.48$ \\
 $2.40$ & $-11.30$  & $-1.20$ & $-4.34$ & $-0.60$ \\
 $3.45$ & $-9.99$   & $-1.76$ & $-4.92$ & $-0.67$ \\
\end{tabular}
\end{ruledtabular}
\end{table}

The MRCI results put forward $K$ as the leading term, whereas DFT yields an even stronger $\Gamma$. A similar discrepancy has been reported for $\alpha$-Li$_2$IrO$_3$~\cite{winter2016,nishimoto2016} and requires further investigation going beyond the scope of our present study. At this point, we only mention that, despite differences on the quantitative level, both DFT and quantum chemistry yield similar pressure evolution of $J$, $K$, and $\Gamma$. These trends, the enhancement of $J$ and $\Gamma$ and the weakening of $K$, are also consistent
with the structural changes reported in Table~\ref{tab:structures}, because the reduction in the Ir--O--Ir angles toward $90^{\circ}$ should indeed 
reduce $K$ \cite{winter2016,nishimoto2016} and enhance $J$ and $\Gamma$~\cite{winter2016}.

\end{document}